\documentclass[fleqn,aps,pra,amsmath,amssymb,preprint,showpacs]{revtex4-1}
\usepackage{amsmath}
\usepackage{graphicx}% Include figure files %[dvipdfmx] for revtex
\usepackage{dcolumn}% Align table columns on decimal point
\usepackage{bm}% bold math
\usepackage{textcomp}
\usepackage{natbib}
\usepackage{enumerate}

\newcommand{\etal}{{\it et al.\ }}

\begin{document}
\title{Precise measurement of hyperfine structure in the $ \rm {3\,S_{1/2}} $ state of $ \rm{^7Li} $} 
\author{Pushpander Kumar}
\author{Vasant Natarajan}
\affiliation{Department of Physics, Indian Institute of Science, Bangalore 560012, India}

\begin{abstract}
We report a precise measurement of hyperfine structure in the $ \rm {3\,S_{1/2}} $ state of the odd isotope of Li, namely $ \rm {^7Li} $. The state is excited from the ground $ \rm {2\,S_{1/2}} $ state (which has the same parity) using two single-photon transitions via the intermediate $ \rm {2\,P_{3/2}} $ state. The value of the hyperfine constant we measure is $ A = 93.095(52)$ MHz, which resolves two discrepant values reported in the literature measured using other techniques. Our value is also consistent with theoretical calculations.
\noindent
%\textbf{Keywords}: Hyperfine constant; Excited state; Lithium.

\end{abstract}

\maketitle

\section{Introduction}

The simple electronic structure of Li lends itself to atomic-structure calculation from first principles \cite{YND08}. However, experimental measurements are complicated by the fact that Li is highly reactive with most transparent materials. This precludes the use of vapor cells (as in the case of other alkali-metal atoms), and the technique of saturated absorption spectroscopy (SAS) to get narrow Doppler-free hyperfine peaks.

The standard way to solve this problem is to collect fluorescence from an atomic beam excited by a perpendicular laser beam. The perpendicularity ensures that the first-order Doppler effect is minimized; but it is not really zero (Doppler free) because of a small misalignment angle from perpendicularity and any divergence of the beam. In fact, the lineshape of each peak is not Lorentzian but Voigt (combination of Lorentzian and Gaussian). However, if one still wants an SAS spectrum with Doppler-free Lorentzian peaks, then the solution is to use an actively pumped stainless steel (SS) chamber with high-enough vapor pressure of Li to get significant absorption \cite{SMN10}.

In this work, we use the atomic-beam technique to measure the hyperfine interval in the $ \rm {3\,S_{1/2}} $ state of $ \rm {^7Li} $. The state is populated using a two-step laser excitation process---the first one is a diode laser at 671 nm to populate intermediate $ \rm {2\,P_{3/2}} $ state, while the second one is a diode laser at 813 nm takes it to the upper $ \rm {3\,S_{1/2}} $ state. The motivation for the measurement is that there are two conflicting high-precision values reported in the literature---one using Stark spectroscopy of Rydberg state reported in 1995 \cite{SIW95}, and the second using two-photon laser spectroscopy reported in 2011 \cite{BNE03}. The value from the former is 189.36(43) MHz, while the value from the latter is 186.212(22) MHz. Our technique is different because it uses two electric dipole allowed single-photon transitions, and does not rely on locking to the second laser. Our value of 186.19(10) MHz is consistent with the more recent measurement reported in \cite{BNE03}. This value is also in good agreement with theoretical calculations \cite{YMD96,GFJ01}.

\section{Experimental details}

\begin{figure}
	\centering
	\includegraphics[width=.5\textwidth]{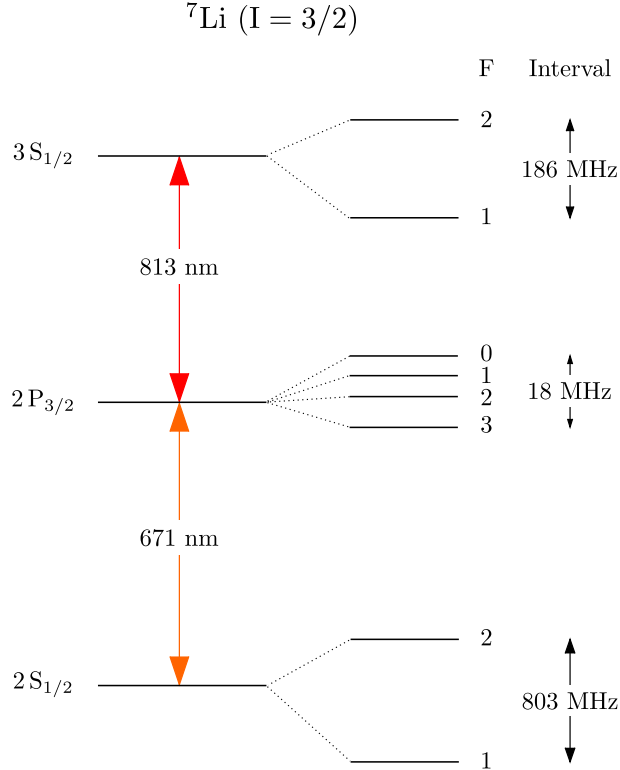}
	\caption{(Color online) Relevant low-lying energy levels of $ \rm {^7Li} $ (not to scale).}
	\label{Li_levels}
\end{figure}

The relevant low-lying energy levels of Li are shown in Fig.~\ref{Li_levels}. The lower transition from the ground state is the $ \rm {2\,S_{1/2} \rightarrow 2\,P_{3/2}} $ transition at 671 nm. The upper transition from the intermediate state is the $  \rm {2\,P_{3/2} \rightarrow 3\,S_{1/2}} $ transition at 813 nm. Both transitions are strong because they are electric dipole (E1) allowed. Hence they can be driven using low-intensity laser beams.

\begin{figure}
	\centering
	\includegraphics[width=.7\textwidth]{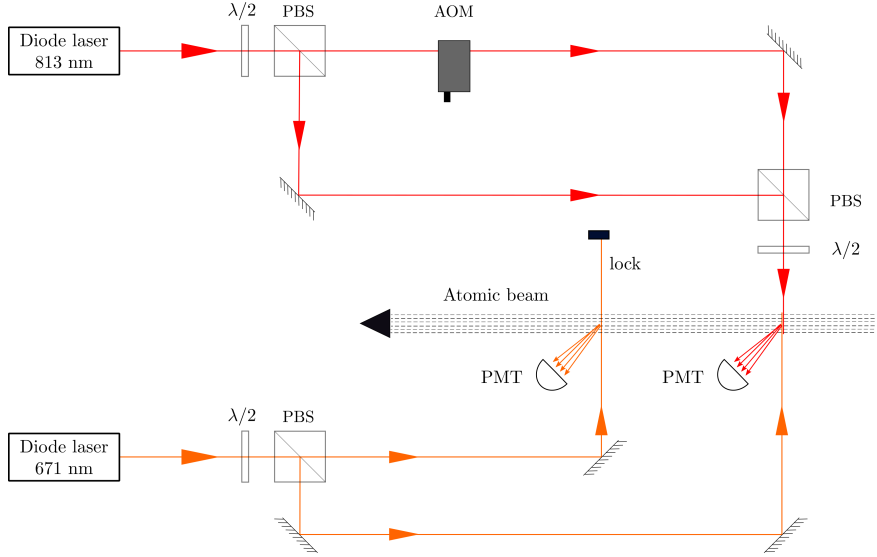}
	\caption{(Color online) Experimental schematic. Figure key: $ \lambda/2 $ -- halfwave retardation plate; PBS -- polarizing beam splitter; AOM -- acousto-optic modulator; PMT -- photomultiplier tube.}
	\label{Li_schematic}
\end{figure}

The experimental setup is shown schematically in Fig.~\ref{Li_schematic}. The laser beams driving the two transitions are derived from feedback-stabilized diode laser systems, as described in Ref.~\cite{MRS15}. The one at 671 nm uses a grating with 2400 lines/mm, while the one at 813 nm uses a grating with 1800 lines/mm.

The laser beam at 671 nm has an elliptic shape, with $ 1/e^2 $ diameter of 1 mm x 4 mm. Its power is controlled using a halfwave ($ \lambda/2 $) retardation plate followed by a polarizing beam splitter cube (PBS). The laser frequency is locked using modulation at 30 kHz of the injection current into the laser diode. As seen from Fig.~\ref{Li_levels}, the hyperfine levels in the $ \rm {D_2} $ line ($ \rm 2\,S_{1/2} \rightarrow 2\,P_{3/2} $ transition) are too closely spaced to be resolved completely. Therefore, the laser is locked to the unresolved $ \rm {F = 2 \rightarrow F^{\prime}} $ peak in the $ \rm {D_2} $ line---the exact lock point is not important as long as transitions to the upper state are allowed. Both $ \rm {1 \rightarrow F^{\prime}} $ and $ \rm {2 \rightarrow F^{\prime}} $ will work; however, $ \rm {2\rightarrow F^{\prime}} $ is chosen because the peak is more prominent.

The laser beam at 813 nm also has an elliptic shape with $ 1/e^2 $ diameter of 1.5 mm x 4 mm. Both the unshifted and AOM-shifted 813 nm beams are used for the experiment. As seen in the figure, the two beams are separated and combined using PBSs. The polarization of the combined beam is adjusted using a $ \lambda/2 $ plate. The combined beam counter-propagates with the locked 671 nm beam for the two-step excitation to the $ \rm {3\,S_{1/2}} $ state. The polarization of the combined 813 nm beam is adjusted to get significant heights for both unshifted and AOM-shifted beams.

All the required spectroscopy experiments are done by having an atomic beam inside an ultra-high vacuum (UHV) system, maintained at a pressure below $ 10^{-7} $ torr using a 40 l/s ion pump. The Li source consists of an SS vial containing a small ingot of unenriched Li. The vial is resistively heated to a temperature of about 200\textdegree C. When heated, the source produces an atomic beam containing both stable isotopes of Li, namely $ \rm {^6Li} $ and $ \rm {^7Li} $. The atomic beam is mechanically collimated with a divergence angle of 0.1 mrad using apertures. The pressure rises by 2 orders-of-magnitude when the source is turned on. The two laser beams intersect the atomic beam at right angles, which as mentioned before minimizes the first-order Doppler shift. The fluorescence signals are collected by photomultiplier tubes (PMTs).

\section{Results and discussion}
\subsection{Experimental results}

\begin{figure}
	\centering
	\includegraphics[width=.7\textwidth]{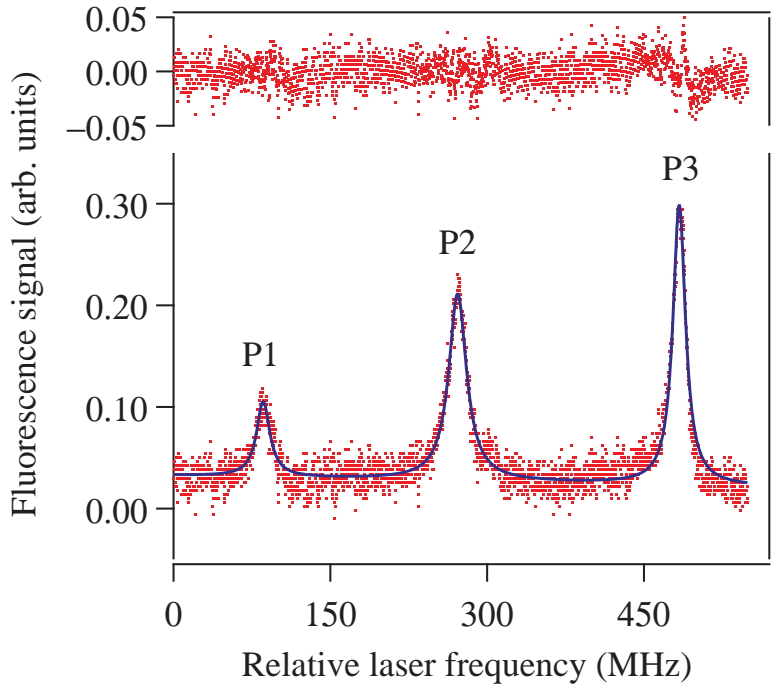}
	\caption{(Color online) Fluorescence signal from the $ \rm {3\,S_{1/2}  \rightarrow 2\,P_{3/2}} $ spontaneous decay of the upper state. P1 corresponds to the $ F = 2 $ level; P2 corresponds to the $ F = 1 $ level; and P3 corresponds to the $ F = 1 $ level again but shifted by the AOM frequency. The solid line is a Lorentzian fit to the 3 peaks with the fit residuals shown on top. }
	\label{Li_spectrum}
\end{figure}

A typical spectrum in $ \rm {^7Li} $ used for the experiment is shown in Fig.~\ref{Li_spectrum}. The fluorescence signal obtained from decay of the $ \rm {3\,S_{1/2}} $ state to the intermediate $ \rm {2\,P_{3/2}} $ state is plotted as a function of laser frequency. The AOM shift is adjusted for 212 MHz. The first peak (P1) corresponds to the $ F = 2 $ level of the $ \rm {3\,S_{1/2}} $ state; the second peak (P2) corresponds to the $F = 1$ level of the $ \rm {3\,S_{1/2}} $ state; and the third peak (P3) is the second peak along with the AOM shift. The solid line is a multipeak Lorentzian fit to the 3 peaks. Even though, as mentioned in the introduction, the lineshape of each peak is a Voigt function, we have used a Lorentzian function because it fits the data quite well, as seen from the featureless residuals shown on top. In addition, the error (which depends on the signal-to-noise ratio) is well-defined for a Lorentzian function. The linewidth of each peak is 15--20 MHz, which is larger than the natural linewidth of 5.25 MHz \cite{LIN77}. The increase in linewidth arises for the following reasons.
\begin{enumerate}[i)]
	\item Small misalignment angle from perpendicularity of the laser and atomic beams.
	\item Residual divergence of the atomic beam.
	\item Closely spaced (less than the natural linewidth) hyperfine levels of the $ \rm {2\,P_{3/2}} $ state.
\end{enumerate}

Since the hyperfine interval to be measured is near 190 MHz, the AOM shift is varied from 160 to 212 MHz, in steps of 2 MHz. At each value of AOM shift, a spectrum of the kind shown in Fig.~\ref{Li_spectrum} is recorded. A multipeak fit with Lorentzian lineshape for the 3 peaks yields each peak's location and error in the location. The hyperfine separation (HFS) is the difference in location between peaks P1 and P2, while the laser scan axis is scaled by the known AOM separation between peaks P2 and P3. The quantity
\begin{equation*}
\begin{aligned}
\delta &= {\rm HFS - AOM} \\
&= {\rm (P3-P1) - 2\, AOM}
\end{aligned}
\end{equation*}		 
has a zero crossing when the AOM frequency is equal to the HFS. The above expression shows that the error in $ \delta $ is equal to the sum of the errors in P1 and P3.

\begin{figure}
	\centering
	\includegraphics[width=.7\textwidth]{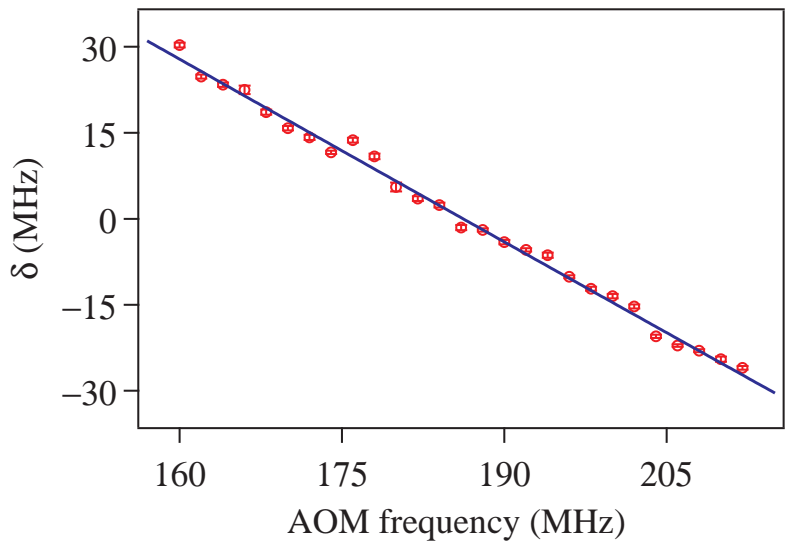}
	\caption{(Color online) Peak separation $ \delta $ (as defined in the text) plotted as a function of AOM frequency. The solid line is a weighted second-order polynomial fit, weighted by the error bar for each point. The error bar for each point is smaller than the symbol, and not seen clearly.}
	\label{secondfit}
\end{figure}

The quantity $ \delta $ as a function of AOM frequency is shown in Fig.~\ref{secondfit}. Each value also has an error bar as determined above. The solid line is a weighted second-order polynomial fit, weighted by the error bar for each point. First order (or linear) is not correct because the laser scan axis is inherently non-linear, varying as the sine of the grating angle. We have also verified that the zero crossing of the fit remains unchanged when we use higher-order polynomials. The zero crossing of the fit along with its error yields the HFS as 186.19(10) MHz. Since the hyperfine interval is related to the hyperfine constant as $ 2A $, the measured value of the constant is $ A = 93.095 \pm  0.050 $ MHz, where the error is the statistical error in the curve fit.

Scanning the laser to get the entire spectrum has many advantages compared to the other technique that we have developed where the AOM frequency is locked to a hyperfine peak \cite{DAN08}. The main advantage is that the technique avoids servo-loop errors. Another advantage is that the measured interval is independent of scaling of the laser scan axis. Any such rescaling will change the $ y $-axis of Fig.~\ref{secondfit}, but not the zero crossing.

\subsection{Error analysis}
The different sources error in the measurement, and our estimated value for each, are listed below.
\begin{enumerate}
	\item Statistical error in the curve fit -- 50 kHz.
	\item AC Stark shift, which causes the lineshape to deviate from Lorentzian -- 5 kHz.
	\item Optical pumping into Zeeman (magnetic) sublevels in the presence of stray magnetic fields -- 10 kHz.
	\item Velocity redistribution due to radiation pressure -- 5 kHz.
	\item Collisional shift -- 1 kHz.
	\item AOM frequency timebase error -- 0.5 kHz.
\end{enumerate}
Adding the above sources of error in quadrature yields the final error in the measurement as 51.5 kHz. Thus the value of the hyperfine constant measured in this work is 
\[
A = 93.095 \pm 0.052 \ \ \text{MHz}
\]

\subsection{Comparison to previous values}
\begin{figure}
	\centering
	\includegraphics[width=.7\textwidth]{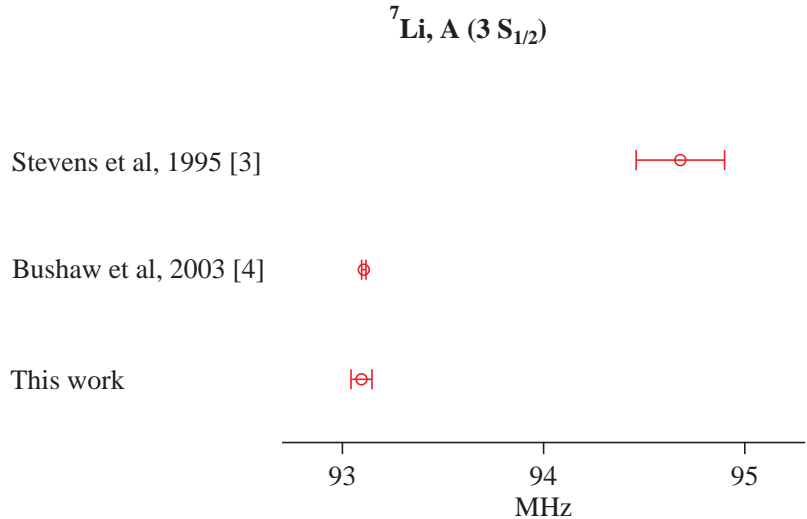}
	\caption{(Color online) Hyperfine constant $ A $ in the $ \rm {3\,S_{1/2}} $ state of $ \rm {^7Li} $ measured in this work compared to earlier measurements.}
	\label{scatter_plot}
\end{figure}

Fig.~\ref{scatter_plot} shows a comparison of our measurement to two previous experimental values. It is clear that our present measurement is consistent with the work of Bushaw \etal \cite{BNE03}, but quite inconsistent with the work of Stevens \etal \cite{SIW95}. Our value is also consistent with theoretical calculations, done with both Hylleraas variational \cite{YMD96} and multiconfiguration Hartree-Fock methods \cite{GFJ01}.

\section{Conclusions}

In summary, we have measured the hyperfine constant in the $ {\rm 3\,S_{1/2}} $ state of $ {\rm ^7Li} $. The state is populated using two single-photon transitions via the intermediate $ \rm {2\,P_{3/2}} $ state. Both transitions are excited using diode lasers. This method is different from previous techniques used in Refs.~\cite{SIW95} and \cite {BNE03}, which report discrepant values for the  hyperfine constant. Our value of $ A = 93.095(52) $ MHz is consistent with the more recent measurement in Ref.~\cite{BNE03}, which uses two-photon spectroscopy for excitation from the ground state. Our value is also consistent with theoretical calculations.

\section*{Acknowledgments}
This work was supported by the Department of Science and Technology, India. The authors thank S Raghuveer for help with the manuscript preparation.  P K acknowledges financial support from the Council of Scientific and Industrial Research, India;

%\bibliography{eitrefs}

%

\end{document}